# INTERPLAY OF h/e AND h/2e OSCILLATIONS IN GATE-CONTROLLED AHARONOV-BOHM RINGS

I.A. Shelykh[1,2], N.T. Bagraev[3], N.G. Galkin[4] and L.E. Klyachkin[3]

[1]LASMEA UMR 6602, Université Blaise Pascal,24, av. des Landais, 63177, Aubiere, France
[2]St. Petersburg State Polytechnical University, 195251, St. Petersburg, Russia
[3]1A.F.Ioffe Physicotechnical Institute, 194021, St. Petersburg, Russia
[4]Mordovian State University, 430000, Saransk, Russia

The spin-interference that is caused by the Rashba spin-orbit interaction in a gate-controlled Aharonov-Bohm ring is studied by the analysis of the conductance oscillations as a function of both the gate voltage and magnetic field. The scattering matrix approach is used to reveal the effect of the quantum scatterers connected to two one-dimensional leads on the phase of the transmission and reflection amplitudes. The variations of the transmission and reflection amplitudes that are caused by the quantum scatterers for the particles moving inside and outside rings are shown to define a parity of the $h/e$ and $h/2e$ conductance oscillations.

The spin-correlated transport in low-dimensional systems was in focus of both theoretical and experimental activity in the last decade [1-3]. The studies of the spin-orbit interaction (SOI) that results from both the crystal and the structure inversion asymmetry in mesoscopic nanostructures have specifically attracted much of the efforts [4-16]. The first mechanism called the Dresselhaus SOI gives rise to the energy separation between the spin bands that is proportional to the cube of the particle wave number, $k^3$ [17]. The Dresselhaus SOI becomes dominant in bulk structures, whereas the second mechanism called the Rashba SOI appears to lift the spin degeneracy of the wave vector parallel to the quantum well (QW) thereby leading to the spin splitting at the Fermi energy that is linear on $k$ [18]. The Rashba SOI has been found to dominate over the Dresselhaus SOI in the Si-MOSFET [19] as well as in InAs/GaSb, AlSb/InAs and GaAs/GaAlAs heterostructures [9, 13-16, 20-22] because of the macroscopic potentials along the interface, which result in the electric field perpendicular to the two-dimensional electron/hole gas.

The Rashba SOI parameter $\alpha$ dependent linearly on the external electric field is of importance to be tuned by varying the gate voltage [9, 13-15]. These variations of the spin splitting at the Fermi energy cause the spin interference effects that have been revealed by beating in

Shubnikov-de Haas oscillations [23] and the transition from the weak antilocalization (WAL) to the weak localization, which was observed as the crossover from positive to negative magnetoresistance near a zero-magnetic field [13-15]. Besides, the spin interference appeared to be a basis of spintronic devices such the spin field effect transistor [1] and the spin-interference device that is able to demonstrate the spin FET characteristics even without ferromagnetic electrodes and external magnetic field [15, 24]. This device shown schematically in Fig. 1a represents the Aharonov-Bohm (AB) ring covered by the gate electrode, which in addition to the geometrical Berry phase [7,8] provides the phase shift between the transmission amplitudes for the particles moving in the clockwise and anticlockwise direction, $\tau_+ - \tau_-$:

$$\tau_\pm = \exp[i(\pi k_\pm a \pm \frac{e\Phi}{2\hbar c})] \quad (1)$$

where $\Phi = \pi a^2 B$; $a$ is the radius of the AB ring. In the absence of the external magnetic field, the phase shift results from the difference in the magnitude of the wave numbers $k_+$ and $k_-$ that is due to the spin orientation relative to the direction of the effective magnetic field $\mathbf{B}_{eff} = \frac{\alpha}{g_B \mu_B}[\mathbf{k} \times \mathbf{e}_z]$, which is caused by the Rashba SOI:

$$k_+ = \frac{m\alpha}{\hbar^2} + \sqrt{\left(\frac{m}{\hbar^2}\right)\left(\frac{m\alpha^2}{\hbar^2} + 2E\right)} \quad (2)$$

$$k_- = -\frac{m\alpha}{\hbar^2} + \sqrt{\left(\frac{m}{\hbar^2}\right)\left(\frac{m\alpha^2}{\hbar^2} + 2E\right)} \quad (3)$$

where $m$ is the effective mass and $E$ is the energy of a particle.

Owing to such a gate-controlled Rashba SOI, the modulation of the conductance was manifested not only in the Aharonov-Bohm oscillations as a function of magnetic field, but in the Aharonov-Casher oscillations as a function of the gate voltage that are caused by the variations of the Rashba SOI parameter $\alpha$ in the absence of external magnetic field [15,24]. Aside from the $h/e$ oscillations that are usually detected in the double-split AB rings, however, the $h/2e$ component, which seems to be the Altshuler-Aronov-Spivak (AAS) oscillations due to the round trip interference in the ring [25], has been surprisingly observed. This result addresses the question whether the mechanism of the AAS oscillations is of sufficiently general character to be applied to the investigation of the spin-interference due to the Rashba SOI in the gate-controlled AB rings. Here we show that the quantum scatterers connected to one-dimensional leads are able to effect on the phase of the transmission and reflection amplitudes for the particles moving inside and outside rings thereby defining a parity of the $h/e$ and $h/2e$ conductance oscillations.

These quantum scatterers that represent the quantum point contacts (QPCs) being the coupling agents between the AB ring and the leads are presumed to be identical and symmetrical with respect to the particles incident from the left and the right. Further treatment of the spin-dependent transport is conducted within the approximation of a weak magnetic field, which produces only the foregoing phase shift in the absence of any spin subbands created by the Zeeman splitting in both the leads and the AB ring. Besides, the spin projection is suggested to be conserved during the scattering of a particle by the QPC. Therefore the transport of the particles with opposite spin projections appears to be treated separately.

These suggestions allow to introduce the scattering matrix $S_1$ and $S_2$ that are independent of the spin projections and able to relate the outgoing current amplitudes $a_2, d_2$ to the incoming current amplitudes $a_1, d_1$ on the QPCs (Fig. 1b) [26-28]. Since the time-reversal invariant case is only taken into account and the QPCs are totally equivalent, the amplitudes of the waves inside the AB ring $a_j, b_j$ and $c_j$, $j=1,2$, which are shown in Fig. 1b, are connected by the following set of the six linear equations:

$$\begin{pmatrix} b_1 \\ a_2 \\ c_1 \end{pmatrix} = \begin{pmatrix} r & \varepsilon & t \\ \varepsilon & \sigma & \varepsilon \\ t & \varepsilon & r \end{pmatrix} \cdot \begin{pmatrix} b_2 \tau_+ \\ 1 \\ c_2 \tau_- \end{pmatrix} \quad (4)$$

$$\begin{pmatrix} b_2 \\ d_2 \\ c_2 \end{pmatrix} = \begin{pmatrix} r & \varepsilon & t \\ \varepsilon & \sigma & \varepsilon \\ t & \varepsilon & r \end{pmatrix} \cdot \begin{pmatrix} b_1 \tau_+ \\ 0 \\ c_1 \tau_- \end{pmatrix} \quad (5)$$

Here $r$ is the reflection amplitude from a QPC to itself, $t$ is the transmission amplitude from a QPC to another QPC, $\sigma$ is the reflection amplitude from the AB ring to itself, $\varepsilon$ is the transmission amplitude from a lead to the AB ring or from the AB ring to a lead. The scattering amplitudes $r, t, \sigma$ and $\varepsilon$ are assumed to be real numbers, $\varepsilon \neq 0$, and to be independent of the energy $E$. Besides, the number of independent matrix elements has to be reduced, because the scattering matrix is unitary owing to the conservation of the flux. Therefore the parametrisation of the scattering matrix performed following by Buttiker et al [28] demands

$$r = \frac{\lambda_1 + \lambda_2\sqrt{1-2\varepsilon^2}}{2} \qquad (6)$$

$$t = \frac{-\lambda_1 + \lambda_2\sqrt{1-2\varepsilon^2}}{2} \qquad (7)$$

$$\sigma = \lambda_2\sqrt{1-2\varepsilon^2} \qquad (8)$$

where $\lambda_{1,2} = \pm 1$. Thus, the effect of the QPCs on the scattering of a particle in the AB ring seems to be defined by the only parameter $\varepsilon$, $-1/\sqrt{2} < \varepsilon < 1/\sqrt{2}$.

The equations (2)-(8) lead to the expression for the reflection amplitude of the device

$$B = a_2 = \sigma + (\varepsilon\tau_-)b_2 + (\varepsilon\tau_+)c_2 \qquad (9)$$

where

$$b_2 = \frac{\varepsilon\left[\left(1-(t\tau_+)^2 - r^2\tau_+\tau_-\right)(r\tau_+ + t\tau_-) + rt\tau_+(\tau_+ + \tau_-)(r\tau_- + t\tau_+)\right]}{\left[1-(t\tau_+)^2 - r^2\tau_+\tau_-\right]\cdot\left[1-(t\tau_-)^2 - r^2\tau_+\tau_-\right] - (rt)^2\tau_+\tau_-(\tau_+ + \tau_-)^2} \qquad (10)$$

$$c_2 = \frac{\varepsilon\left[\left(1-(t\tau_-)^2 - t^2\tau_+\tau_-\right)(r\tau_- + t\tau_+) + rt\tau_-(\tau_+ + \tau_-)(r\tau_+ + t\tau_-)\right]}{\left[1-(t\tau_+)^2 - r^2\tau_+\tau_-\right]\cdot\left[1-(t\tau_-)^2 - r^2\tau_+\tau_-\right] - (rt)^2\tau_+\tau_-(\tau_+ + \tau_-)^2} \qquad 11)$$

Finally, the formulae permit to calculate the conductance of the AB ring:

$$G = 2\frac{e^2}{h}\left(1-|B|^2\right) \qquad (12),$$

which is dependent on the energy of a particle, the value of external magnetic field, and the Rashba SOI parameter $\alpha$. These dependencies are difficult to be present analytically, but the conductance of the AB ring in the absence of the effect of the QPCs on the scattering of particles, $\varepsilon = \pm 1/\sqrt{2}$, is specifically of interest. Using the simple algebra, the expression for the reflection amplitude has to find in this particular case:

$$B(\alpha, E, \Phi) = \frac{\sin^2\left(\frac{\pi m \alpha a}{\hbar^2} + \frac{e\Phi}{2hc}\right)}{1 - \exp\left(2\pi i a \sqrt{\left(\frac{m}{\hbar^2}\right)\left(\frac{m\alpha^2}{\hbar^2} + 2E\right)}\right)\cos^2\left(\frac{\pi m \alpha a}{\hbar^2} + \frac{e\Phi}{2hc}\right)} \quad (13)$$

This expression predicts the quasiperiodical and asymmetrical oscillations of a conductance as a function of the Rashba SOI parameter $\alpha$ as opposed to the straightforward formula,

$$G = \frac{e^2}{h}\left[1 + \cos(2\pi a \frac{\alpha m}{\hbar^2})\right] \quad (14),$$

which has been deduced as the first order approximation [24] (Fig. 2a). The AC oscillations with similar quasiperiodicity appeared to be revealed by varying the gate voltage applied to the AB ring inserted in the InGaAs/InAlAs heterostructure [15]. The AB oscillations calculated using Eq. (13) are of interest to exhibit as before the standard $h/e$ periodicity shown in Fig. 3a thereby defining the broad periodical plateau of the $2e^2/h$ conductance as a function of $\alpha$ and $B$ (Fig. 4a). The results of these calculations are however contrast to the studies of the AB oscillations in the same InGaAs/InAlAs heterostructure, which demonstrate both $h/e$ and $h/2e$ harmonics as a function of the external magnetic field value [15]. Therefore, using the Eqs. (9)-(11) for the numerical conductance calculations of the gate-controlled AB rings in the presence of the scattering by the quantum scatterers, $-1/\sqrt{2} < \varepsilon < 1/\sqrt{2}$, is the most promise for the spin-interference devices, because the part of the gate and/or drain-source voltage appears to be applied presumbly at QPCs giving rise to the changes of the $\varepsilon$ value.

The results of numerical calculations shown in Figs. 2b, 2c and 3b, 3c reveal the progressive enhancement of the quasiperiodicity of both AC and AB oscillations with reduction in the $\varepsilon$ value, which plot specifically as a gradual transformation of the $2e^2/h$ plateaus in the $2e^2/h$ pronounced peaks as a function of $\alpha$ and $B$ (Figs. 4b and 4c). The $h/2e$ harmonic is of importance to dominate over the $h/e$ harmonic already in a slight effect of the quantum scatterers on the conductance of the AB rings (Figs. 2b and 3b). Thus, the scattering of the particles moving in the clockwise and anticlockwise direction by QPCs seem to enhance the variations of the amplitudes and the phase of the AC and AB oscillations in the gate-controlled AB rings [15, 24] and to influence on the amplitudes of the AB oscillations in the double-split AB rings [29]. Further experimental investigations of the mechanisms of the AAS oscillations revealed by the spin-interference due to the Rashba SOI in the gate-controlled AB rings are underway.

In summary, we have applied the scattering matrix formalism to the analysis of the spin-interference that is caused by the Rashba spin-orbit interaction in a gate-controlled AB rings. The effect of the quantum scatterers that represent the quantum point contacts connecting the AB ring to two one-dimensional leads on the phase of the transmission and reflection amplitudes of the AC and AB and oscillations has been revealed by varying the value of the gate voltage and external magnetic field. The variations of the transmission and reflection amplitudes that are caused by the quantum scatterers for the particles moving inside and outside rings have been shown to define a parity of the $h/e$ and $h/2e$ conductance oscillations.

This work has been supported by the Marie Curie project "Clermont 2" contract number MRTN-CT-2003-503677.

**References**

1. S. Datta and B. Das, Appl. Phys. Lett. **56**, 665 (1990).
2. S.A. Wolf *et al.*, Science **294**, 1488 (2001).
3. D.D. Awschalom *et al.*, *Semiconductor Spintronics and Quantum Computation*, Springer-Verlag, 2002.
4. B.L. Altshuler and A.G. Aronov, *in Electron-Electron Interactions in Disordered Systems*, edited by A.L. Efros and M. Pollak, North Holland, Amsterdam, 1985, p.1.
5. S. Hikami, A.I. Larkin, and Y. Nagaoka, Prog. Theor. Phys. **63**, 707 (1980).
6. G. Bergmann, Sol. St. Commun. **42**, 815 (1982).
7. A.G. Aronov and Y.B. Lyanda-Geller, Phys. Rev. Lett. **70**, 343 (1993).
8. A.F. Morpurgo *et al.*, Phys. Rev. Lett. **80**, 1050 (1998).
9. W. Knap et al., Phys. Rev **B 53**, 3912 (1996).
10. A.V. Moroz and C.H.W. Barnes, Phys. Rev. **B 60**, 14272 (1999).
11. F. Mierles and G. Kirczenov, Phys. Rev. **B 64**, 024426 (2001).
12. W.D. Oliver, G.Feve and Y. Yamamoto, Journal of Superconductivity **16**, 719 (2003).
13. J.B. Miller *et al.*, Phys. Rev. Lett. **90**, 076807 (2003).
14. S.A. Studenikin *et al.*, Phys. Rev. **B 68**, 035317 (2003).
15. J. Nitta and T. Koga, Journal of Superconductivity **16**, 689 (2003).
16. A. Ghosh *et al.*, Phys. Rev. Lett. **92**, 116601 (2004).
17. G. Dresselhaus, Phys. Rev. **100**, 580 (1955).
18. E.I. Rashba, Sov. Phys. Solid State **2**, 1109 (1960); Yu. A. Bychkov and E.I. Rashba, Sov. Phys. JETP Lett. **39**, 78 (1984); Yu. A. Bychkov and E.I. Rashba, J.Phys. **C17**, 6039 (1984).
19. S. I. Dorozhkin and E.B. Ol'shanetskii, Sov. Phys. JETP Lett. 46, 502 (1987).
20. J. Luo *et al.*, Phys. Rev. **B 41**, 7685 (1990).

21. G.L. Chen *et al.*, Phys. Rev. **B 47**, 4084 (1993).
22. T. Hassenkam *et al.*, Phys. Rev. **B 55**, 9298 (1997).
23. V.M. Pudalov *et al.*, Phys. Rev. Lett. **88**, 196404 (2002).
24. J. Nitta, F.E. Meijer, and H. Takayanagi, Appl. Phys. Lett. **75**, 695 (1999).
25. B.L. Altshuler, A.G. Aronov, and B.Z. Spivak, Sov. Phys. JETP Lett. **33**, 94 (1981).
26. T. Taniguchi and M. Buttiker, Phys. Rev. **B 60**, 13814 (1999).
27. H.L. Endquist and P.L. Anderson, Phys. Rev. **B 24**, 1151 (1981).
28. M. Buttiker, Y. Ymry, and M. Ya. Azbel, Phys. Rev. **A 30**, 1984 (1984).
29. R. Schuster, E. Buks, M. Heiblum, D. Mahalu, V. Umansky, and H. Shtrikman, Nature **385**, 417 (1997).

CAPTIONS

Fig. 1. Schematic view of a spin-interference device [15, 24] that is based on the AB ring connected with two one-dimensional leads by QPCs and covered by the gate electrode that controls the Rashba SOI.

(a) The spin directions of the clockwise and anticlockwise travelling electronic waves.
(b) The amplitudes of travelling electronic waves.

Fig. 2. The AC oscillations calculated at different values of the transmission amplitude from a lead to the AB ring (from the AB ring to a lead).

(a) $\varepsilon = 0.707$; (b) $\varepsilon = 0.507$; (c) $\varepsilon = 0.307$.

Fig. 3. The AB oscillations calculated at different values of the transmission amplitude from a lead to the AB ring (from the AB ring to a lead).

(a) $\varepsilon = 0.707$; (b) $\varepsilon = 0.507$; (c) $\varepsilon = 0.307$.

Fig. 4. Grayscale plot of $G$ vs $\alpha$ and $B$

(a) $\varepsilon = 0.707$; (b) $\varepsilon = 0.507$; (c) $\varepsilon = 0.307$.

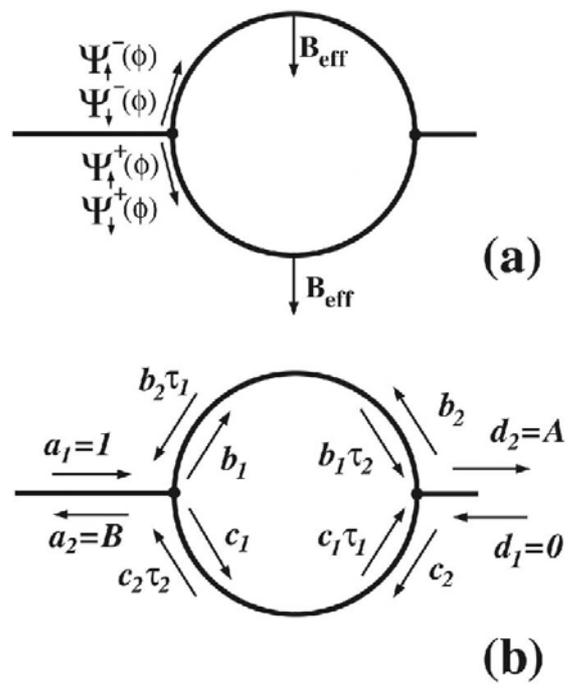

Fig. 1

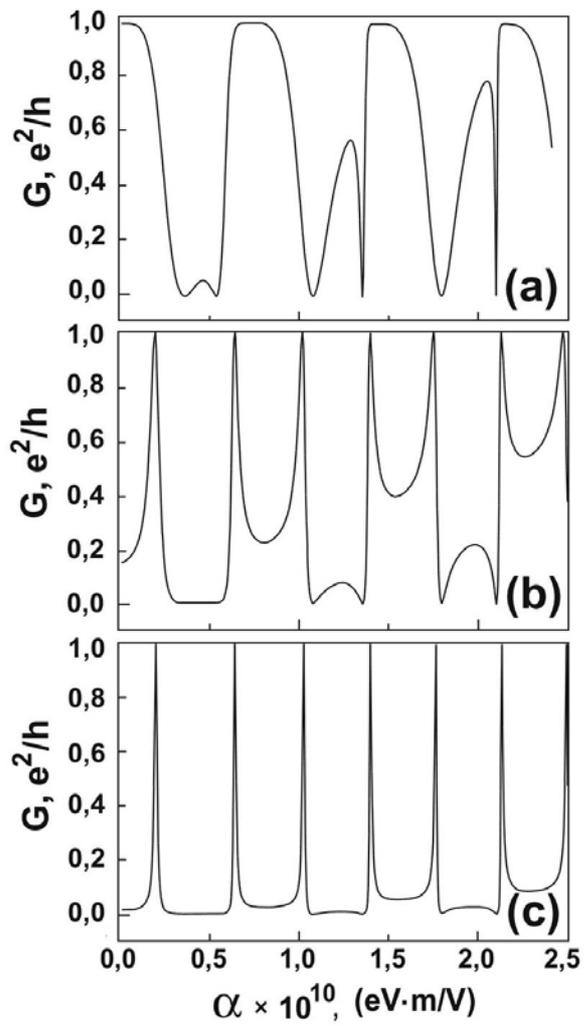

Fig. 2

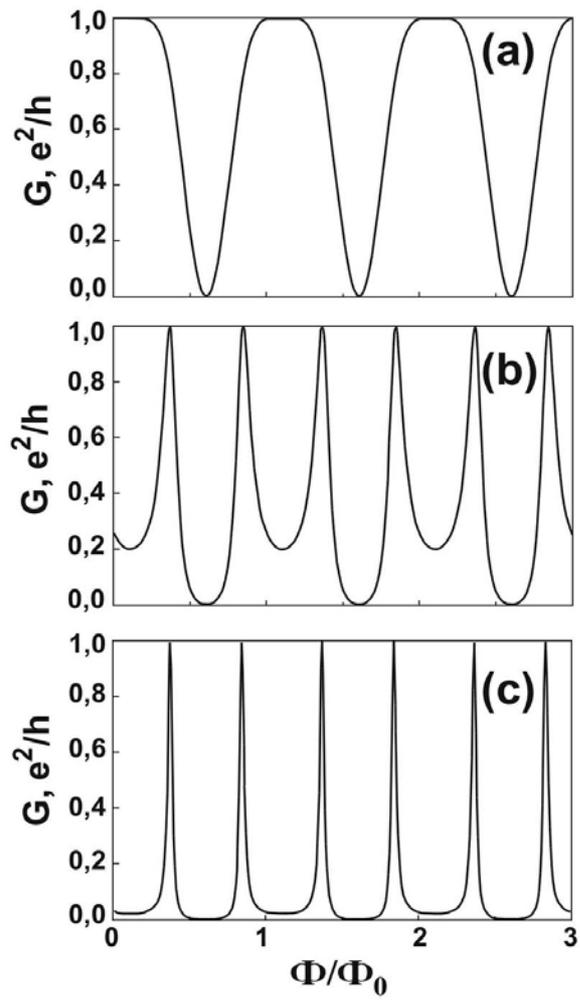

Fig. 3

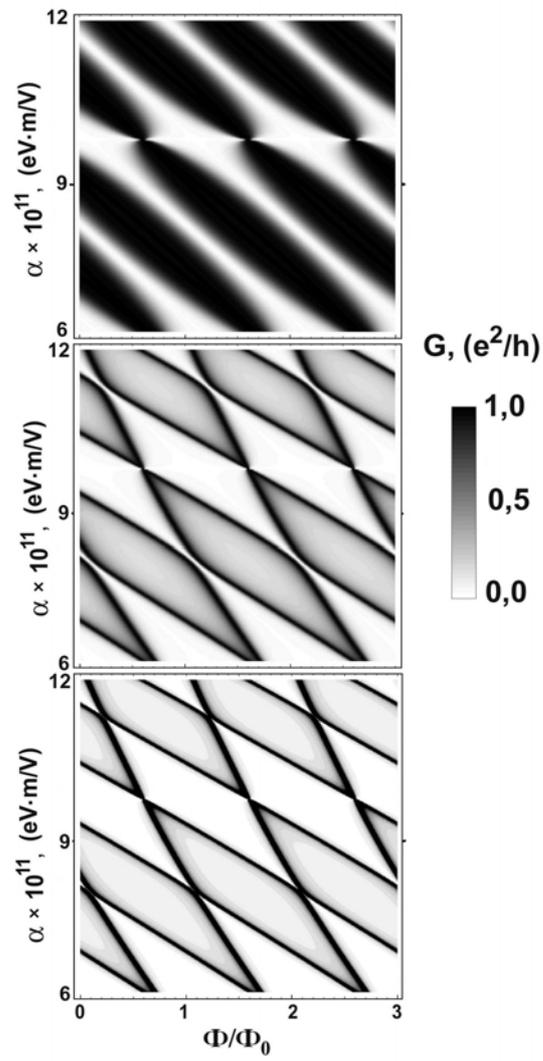

Fig. 4